\documentclass[aps,amsfonts,nofootinbib]{revtex4}
\usepackage{epsfig}
\usepackage{graphicx}
\usepackage{amsmath}
\usepackage{amsbsy}
\usepackage{color}
\usepackage{epsfig}
\usepackage{graphicx}
\usepackage{amsmath}
\usepackage{amssymb}
\usepackage{amsbsy}
\newcommand{\ee}{\end{equation}}
\newcommand{\bb}{\begin{equation}}
\newcommand{\eqb}{\begin{eqnarray}}
\newcommand{\eqf}{\end{eqnarray}}

\def\p{\mathbf{p}}

\newcommand{\1}{{\'{\i}}}

\def\1{\'{\i}}

\begin{document}
\title{Oscillators in a (2+1)-dimensional noncommutative space}

\author{F.\ Vega}

\affiliation{IFLP - CONICET and Departamento de  F\'{\i}sica, Facultad de Ciencias Exactas de la UNLP, \\
C.C.\ 67, (1900) La Plata, Argentina
\\
\\
E-mail: federicogaspar@gmail.com}

\begin{abstract}

We study the Harmonic and Dirac Oscillator problem extended to a three-dimensional noncommutative space where the noncommutativity is induced by the shift of the dynamical variables with generators of $SL(2,\mathbb{R})$ in a unitary irreducible representation considered in reference \cite{Falomir-yo}. This redefinition is interpreted in the framework of the Levi's decomposition of the deformed algebra satisfied by the noncommutative variables.

The Hilbert space gets the structure of a direct product with the representation space as a factor, where there exist operators which realize the algebra of Lorentz transformations. The spectrum of these models are considered in perturbation theory, both for small and large noncommutativity parameters, finding no constraints between coordinates and momenta noncommutativity parameters.

Since the representation space of the unitary irreducible representations  $SL(2,\mathbb{R})$ can be realized in terms of spaces of square-integrable functions, we conclude that these models are equivalent to quantum mechanical models of particles living in a space with an additional compact dimension.

\vskip 0.3cm

\end{abstract}
\date{\today}
\maketitle

\section{Introduction}

In reference \cite{Falomir-yo}, a (2+1)-dimensional model with a kind of nonstandard noncommutativity has been considered, where both coordinates and momenta get deformed commutators. In order to preserve the Lorentz invariance, these deformations of the Heisenberg algebra were taken as proportional to generators of the Lorentz group in some irreducible representation (\emph{irrep}) of the (noncompact) Lie group $SL(2,\mathbb{R})$.

 This deformed algebra can be effectively realized by shifting the ordinary (commutative) dynamical variables by generators of the irreducible representation of $SL(2,\mathbb{R})$ considered. This means that the Hilbert space has the structure of a direct product, where one factor corresponds to the component of the state vectors in the representation space of the \emph{irrep} considered.

As we demand that the noncommuting phase space variables be Hermitian operators, we are constrained to consider unitary \emph{irrep}'s of this group, which are not of finite dimension. Moreover, since the representation space of the unitary \emph{irrep}'s of $SL(2,\mathbb{R})$ can be explicitly realized in terms of spaces of functions defined on the unit circle or analytic functions on the unit open disk (the irreducible representations of $SL(2,\mathbb{R})$ have been extensively discussed in \cite{Bargmann}; see also \cite{Falomir-yo}), the models to be considered turn out to be equivalent to quantum mechanical systems living in a space with an additional (compact) dimension.

\section{Setting of the problem}\label{NCspace}

According to the ideas exposed in  \cite{Falomir-yo}, we consider the modified Heisenberg algebra ($\mathfrak{H}$) of the (Hermitian) dynamical variables given by:
\begin{equation}
\label{d30}
\begin{array}{lll} \displaystyle
\left[{\hat  x}_\mu,{\hat x}_\nu \right]  =- i\theta^2  \epsilon_{\mu \nu \rho} \hat s^\rho \,,  & \quad  &   \left[{\hat   p}_\mu,{\hat   p}_\nu\right]   =   -i   \kappa^2
\epsilon_{\mu \nu \rho} \hat s^\rho\,,
\\  \\ \displaystyle
\left[{\hat  x}_\mu,{\hat p}_\nu\right]  = i(\eta_{\mu \nu}  -  \kappa \theta
\epsilon_{\mu \nu \rho} \hat s^\rho)\,, & \quad &
\left[ {\hat  x}_\mu, \hat s_\nu \right] = - i \theta\epsilon_{\mu \nu \rho}\hat  s^\rho
\,,
\\  \\ \displaystyle
\left[  {\hat  p}_\mu,   \hat s_\nu  \right]  =   -i  \kappa \epsilon_{\mu \nu \rho}\hat  s^\rho
\,, & \quad  &
\left[\hat s_\mu,\hat s_\nu \right] =- i\epsilon_{\mu \nu \rho}\hat s^\rho\,,
\end{array}
\end{equation}
where $\eta:=diag(1,-1,-1)$, the metric tensor in the 2+1-dimensional Minkowski space, $s_\mu\,, \mu=0,1,2$ are the generators of an \emph{irrep} of $SL(2,\mathbb{R})$ and $\theta$ and $\kappa$ play the role of ultraviolet and infrared scales respectively.

We can define the generators of the Lorentz transformations in this 2+1-dimensional noncommutative space independently of the algebra realization. Indeed, if we set: $\hat M_\mu:=\hat L_\mu+\hat s_\mu$, with $\hat L^\lambda:=\frac{1}{2}\, \epsilon^{\lambda\mu\nu}\left[  (\hat x_\nu -\theta \hat s_\nu)(\hat p_\mu -\kappa \hat s_\mu)- (\hat x_\mu -\theta \hat s_\mu) (\hat p_\nu -\kappa \hat s_\nu)\right]$, it can be seen that:
\begin{equation}\label{33}
   \begin{array}{c}\displaystyle
\left[\hat  M_\mu ,\hat  M_\nu \right]= -\imath \epsilon_{\mu \nu \lambda}\hat  M^\lambda\quad
\left[ \hat M_\mu , \hat s_\nu \right]= -\imath \epsilon_{\mu \nu \lambda} \hat s^\lambda\,,
\\ \\ \displaystyle
\left[\hat  M_\mu , \hat{x}_\nu \right]= -\imath \epsilon_{\mu \nu \lambda} \hat{x}^\lambda\,,\quad
\left[\hat  M_\mu , \hat{p}_\nu \right]=- \imath \epsilon_{\mu \nu \lambda} \hat{p}^\lambda\,.
\end{array}
\end{equation}
This equations shows that this kind of non-commutative does not break Lorentz invariance.

In order to have a better insight into the full symmetry of the NC space in question we study the Levi decomposition of this modified Heinsenberg algebra $\mathfrak{H}$. This method provides us a way to define a new and more convenient basis that decompose $\mathfrak{H}$ into a semidirect sum of a solvable ideal and a semisimple subalgebra.

Before giving the statement of the theorem, lets recall some of the definitions that will be used:
\begin{itemize}
\item An algebra $\mathfrak{G}$ is called solvable if there exist a $k\in\mathbb{N}$ such that $\mathfrak{G}^{(k)}=0$ and $\mathfrak{G}^{(k-1)}\neq0$, where $\mathfrak{G}^{(j)}=\left[ \mathfrak{G}^{(j-1)},\mathfrak{G}^{(j-1)}\right]$ and $\mathfrak{G}^{(1)}=\mathfrak{G}$, for any $j\in\mathbb{N}$. 
\item A subalgebra $\mathcal{S}$ is said to be semisimple if it does not contain any solvable ideals exept $\mathbf{1}$.
\item The solvable radical of $\mathfrak{H}$ is by definition the maximal solvable ideal of  $\mathfrak{H}$, which is given by the sum of all solvable ideals of this algebra.
\end{itemize}

Now, we are in conditions to enunciate the Levi decomposition theorem \cite{levi}:
\\ \\
\textit{Given a finite dimensional Lie algebra  $\mathfrak{H}$. If $\mathfrak{H}$ is not solvable then there exist a semisimple subalgebra  $\mathcal{S}$ of  $\mathfrak{H}$ such that  $\mathfrak{H}= \mathcal{S} \oplus SR(\mathfrak{H})$, where $SR(\mathfrak{H})$ is the solvable radical of $\mathfrak{H}$. In this decomposition $\mathcal{S} \sim \mathfrak{H}/SR( \mathfrak{H})$ and we have the following commutation relations:
\begin{equation}
\left[  \mathcal{S};\mathcal{S} \right]  = \mathcal{S}\,,\quad \left[ \mathcal{S};SR(\mathfrak{H})\right]  \subseteq SR(\mathfrak{H})\,,\quad \left[ SR(\mathfrak{H});SR(\mathfrak{H})\right]  \subseteq SR(\mathfrak{H})\,.
\end{equation}}

In order to calculate the $SR(\mathfrak{H})$ we use the following proposition:
\\ \\
\textit{If we are dealing with a finite dimensional Lie algebra, then:
\begin{equation}
SR(\mathfrak{H})=\left\{x\in \mathfrak{H} / \,\,Tr(ad_x\circ ad_y)=0, \forall y \in \left[\mathfrak{H},\mathfrak{H}\right]\right\}\,,
\end{equation}
where $ad_x$ is the adjoint map of $x$, $ad_{x}:=[x, ]$.}

\subsection{Algorithm to compute Levi decomposition}
\begin{itemize}
\item First find a base of the subalgebra $\left[\mathfrak{H},\mathfrak{H}\right]$. In our case, $\left[\mathfrak{H},\mathfrak{H}\right]=span\left\{\mathbf{1},\hat{s}_0, \hat{s}_1, \hat{s}_2\right\}$.
\item For every element of the basis of $\mathfrak{H}$ calculate the adjoint map of that element with respect to all the elements of the basis. Clearly, this is a linear map and has a matrix representation. In our case: $\mathfrak{H}:=span\left\{\hat \xi_i\right\}_{i=0}^{9}$,  where $\hat \xi=\left(\mathbf{1},\hat x_1,\hat x_2,\hat x_3,\hat p_1,\hat p_2,\hat p_3,\hat s_1,\hat s_2,\hat s_3\right)^t$.
\item Compute the Killing map of $\mathfrak{H}$, $\mathcal{K}_{ij}(\mathfrak{H}):=Tr[ad_{\hat \xi_i}\circ ad_{\hat \xi_j}]$. Here $\hat \xi_i$ is an element of the basis of $\mathfrak{H}$ and $\hat \xi_j$ is an element of the basis of $\left[\mathfrak{H},\mathfrak{H}\right]$. Naturally, the composition rule for the adjoint map in the matrix representation reduces to the usual matrix product.
\item Now, we are in conditions to compute a basis for $SR(\mathfrak{H})$. Indeed:
\begin{equation}\label{SL}
x=\sum_{i=0}^9\alpha_i\hat \xi_i\in SR(\mathfrak{H})\iff\sum_{i=0}^9\alpha_i Tr[ad_{\hat \xi_i}\circ ad_{\hat \xi_j}]=0\,,\quad \forall \hat  \xi_j\in\left[\mathfrak{H},\mathfrak{H}\right]
\end{equation}
then all the problem reduces to solve the linear equations for the constant $\alpha_i$.
\end{itemize}

Applying this method we find that $SR(\mathfrak{H})$ is isomorphic to the Heisenberg algebra $\mathcal{H}$, and $S\sim SL(2,\mathcal{R})$, so we can write:
\begin{equation}
\mathfrak{H}:=\mathcal{H}\oplus SL(2,\mathcal{R})\,.
\end{equation}
It is evident from the last equation that we can set as a base of $\mathfrak{H}$ the usual commutative operators of the Heisenberg algebra plus the generators of $SL(2,\mathcal{R})$ in coincidence with the election made in \cite{Falomir-yo}. Other conclusion is that the realization of this algebra is not unique, on the contrary, two representations of $\mathfrak{H}$  will be related through a canonical transformation ($\Omega_{\mathfrak{H}}$) of the algebra. This $\Omega_{\mathfrak{H}}$ can be easily written in terms of the canonical transformations of $\mathcal{H}$ ($\Omega$) and the canonical transformations of the $SL(2,\mathcal{R})$ ($\widetilde{\Omega}$):
\begin{equation}
\Omega_{\mathfrak{H}}=\Omega\otimes\widetilde{\Omega}\,.
\end{equation}

The conditions that determine $\Omega$ and $\widetilde{\Omega}$ are:
\begin{equation}\label{Omega}
\Omega\,\eta_{\mu\nu}\,\Omega^t=\eta_{\mu\nu}\,,\quad \widetilde{\Omega}\,\mathbf{B}^\sigma\widetilde{\Omega}^t=\mathbf{B}^\sigma\widetilde{\Omega}\,,\quad \sigma=0,1,2.
\end{equation}
where we set $(\mathbf{B}^\sigma)_{\mu\nu}:=-(\mathbf{B}^\sigma)_{\nu\mu}=\epsilon_{\mu\nu\rho}\eta^{\rho\sigma}$. Is easy to prove that the second equation in $\eqref{Omega}$ is equivalent to $\widetilde{\Omega}^{t}\mathbf{B}=\mathbf{B}\widetilde{\Omega}$

As in \cite{Falomir-yo} we realize $\mathfrak{H}$ in terms of dynamical variables satisfying the usual Heisenberg algebra, $x^\mu,p_\mu$ with $\mu=0,1,2$, wich corresponds to a kind of \emph{non-Abelian Bopp's shift}:
\begin{equation}\label{shift}
    \hat{x}_\mu\rightarrow x_\mu+\theta s_\mu\,,\quad \hat{p}_\mu\rightarrow p_\mu + \kappa s_\mu\,.
\end{equation}

For this representation $\hat M_\mu$ reduces to the $M_\mu:=L_\mu+s_\mu$, with $L^\lambda:=\frac{1}{2}\, \epsilon^{\lambda\mu \nu }\left(  x_\nu p_\mu - x_\mu p_\nu \right)$.

\medskip

In the following, our strategy to formulate a model in this noncommutative space will be, given a Hamiltonian $H(\mathbf{p},\mathbf{x})$ in the usual (commutative) Minkowski space, generalize it by taking $H(\mathbf{\hat{p}},\mathbf{\hat{x}})$ and then analyze it through the replacements in Eq.\ (\ref{shift}).  While the original paper \cite{Falomir-yo} dealt with the Landau problem, this paper deals with the harmonic oscillator. In the usual commutative space the Landau problem reduces to an harmonic oscillator with an angular momentum term but in this (2+1)-noncommutative space the extension is non trivial because $\hat L_\mu$ has an internal structure given by the generators of $SL(2,R)$.

In particular, as was previously mentioned, we get a space of state vectors which is the direct product of the Hilbert space for the systems in the usual commutative space with the representation space of a unitary \emph{irrep} of $SL(2,\mathbb{R})$. Since these representation spaces can be realized as spaces of square-integrable functions (functions on the unit circle or analytic functions on the open unit disk, according to the particular \emph{irrep} considered) \cite{Bargmann}, these models can also be interpreted as describing particles living in spaces with an additional (compact) spatial dimension.

\section{SCHR\"{O}DINGER Harmonic Oscillator}

\subsection{The usual case}\label{Schrodinger}

In order to set the notation let us first consider the Schr\"{o}dinger Hamiltonian for the harmonic oscillator:
\begin{equation}\label{S1}
    2 M H:=p_i^2+ M^2\omega^2x^2_i\,,
\end{equation}
which commutes with the generator of rotations on the plane, $L_0$.

In terms of creation and annihilation operators (see \cite{Falomir-yo} for the notation) the Hamiltonian $H$ and the angular momentum $L_0$ become diagonal and can be written as
\begin{equation}
H:=\omega(a^{\dagger}a+b^{\dagger}b+1),\quad L_0:=b^{\dagger}b-a^{\dagger}a \,.
\end{equation}
Their eigenvectors for both operators are:
\begin{equation*}
    \left| n_a, n_b \right\rangle := \frac{\left( a^{\dagger} \right)^{n_a}}{\sqrt{n_a!}}\, \frac{\left( b^{\dagger}\right)^{n_b} }{\sqrt{n_b!}}
    \left| 0,0\right\rangle\,, \quad n,N=0,1,2,\dots
\end{equation*}
where $a \left| 0,0\right\rangle=0= b \left| 0,0\right\rangle$, corresponding to the eigenvalues
\begin{equation}
    H \left| n_a, n_b \right\rangle = \omega \left(n_a+n_b+1 \right) \left| n_a, n_b \right\rangle\,,
    \quad  L_0 \left| n_a, n_b \right\rangle = \left(n_b-n_a\right) \left| n_a, n_b \right\rangle\,.
\end{equation}
We call $l:=\left(n_b-n_a\right)$.

\subsection{Extension to the noncommutative space}

The generalization of this system to the noncommutative space is obtained thought the shift defined in Eq.\ (\ref{shift})
\begin{equation}\label{S7}
     2 M \hat{H}:=\hat{p}_i^2+M^2\omega^2\hat{x}^2_i
     = \left( p_i+\kappa s_i \right)^2+M^2\omega^2\left( x_i+\theta s_i \right)^2\,,
\end{equation}
which commutes with $M_0=L_0+s_0$, as can be easily seen.

\medskip

In terms of creation and annihilation operators and the Hermitian generators $s_\mu$, this Hamiltonian reads as
\begin{equation}\label{S8}
    \begin{array}{c} \displaystyle
      2 M \hat{H}
     =2MH+ \sqrt{M\omega}\left(z a^\dagger s_+ + \overline{z}a s_- + z b^{\dagger}s_-+\overline{z} b s_+\right)
     +z\overline{z}\left( {s_0}^2-{\mathbf{s}}^2\right) \,,
    \end{array}
\end{equation}
where $z:= \theta M\omega+\imath \kappa$, $\overline{z}:= \theta M\omega- \imath \kappa$ and we have defined
\begin{equation}\label{gen-sl2r}
    s_\pm:=s_1\pm \imath s_2\,, \quad \mathbf{s}^2:={s_0}^2-{s_1}^2-{s_2}^2\,.
\end{equation}

In \cite{Falomir-yo} one can  find a brief review of the unitary irreducible representations (\emph{irrep}) of $sl(2,\mathbb{R})$. The representation space is generated by the basis of simultaneous eigenvectors of $s_0$ and
$\mathbf{s}^2$,
\begin{equation}\label{16666}
    {\mathbf{s}^2} \left|\lambda,m\right\rangle = \lambda \left|\lambda,m\right\rangle\,, \quad
    s_0 \left|\lambda,m\right\rangle = m \left|\lambda,m\right\rangle\,,
\end{equation}
where $\lambda$ and $m$ are real numbers. We also have
\begin{equation}\label{S12}
    s_\pm \left| \lambda, m \right\rangle = \sqrt{m(m\pm 1)-\lambda}\,   \left| \lambda, m\pm 1 \right\rangle\,.
\end{equation}

The \emph{discrete classes} of unitary \emph{irrep}'s \cite{Bargmann,Falomir-yo} correspond to $\lambda\geq-1/4$, for which this parameter takes discrete values, $\lambda=k(k-1)$ with $k=\frac N 2, N\in \mathbb{N}$, and either $m=k, k+1,k+2, \cdots$ or $m=-k, -k-1,-k-2, \cdots$ On the other hand, for the \emph{continuous classes} of unitary \emph{irrep}'s \cite{Bargmann,Falomir-yo}, $\lambda$ takes any real value less than $-1/4$ and $m$ takes either all the integer or all the half-integer values.

The Hilbert space is then the linear span of the vectors of the form
\begin{equation}\label{S9}
    \left| n_a,n_b; \lambda,m \right\rangle:=\left|n_a,n_b\right\rangle\otimes\left| \lambda, m \right\rangle \,,
\end{equation}
which are simultaneously eigenvectors of $H$, $L_0$, $\mathbf{s}^2$ and $s_0$, normalized so as to satisfy
\begin{equation}\label{S10}
   \left\langle n_a,n_b; \lambda,m |  n_a',n_b'; \lambda,m' \right\rangle = \delta_{n_a,n_a'}\delta_{n_b,n_b'}\delta_{m,m'}\,.
\end{equation}

We also have that $\left[ \hat{H} , M_0 \right]=0$, where $M_0=L_0+s_0$ has eigenvalues $j=l+m=n_b-n_a+m$, integer or half-integer according to the \emph{irrep} of $SL(2,\mathbb{R})$ considered. Then, for given values of $\lambda$ and $j$, we can give the following development for the $\hat{H}$'s eigenvectors,
\begin{equation}\label{S13}
    \left| \psi_{E,j }\right\rangle = \sum_{n_b-n_a+m=j} C_{n_a,n_b,m}   \left| n_a,n_b ; \lambda,m \right\rangle\,.
\end{equation}

\medskip

From Eq.\ (\ref{S8}), one straightforwardly gets the linear recursion relation for the coefficients
\begin{equation}\label{S14}
  \begin{array}{c} \displaystyle
     \left\langle n_a,n_b ; \lambda,m  \right| 2M (\hat{H} -E)  \left| \psi_{E,j,n_b }\right\rangle=
     \left\{ 2M\omega(n_a+n_b+1)-2M(E-\kappa m)+\bar{z} z \left( m^2-\lambda\right)  \right\} C_{n_a,n_b,m}+
     \\ \\ \displaystyle
     +z \sqrt{M\omega} \sqrt{n_a+1}\sqrt{m(m+1)-\lambda}\,  C_{n_a+1,n_b,m+1} +
      \bar{z}  \sqrt{M\omega}\sqrt{n_a}\sqrt{(m-1)m-\lambda}\,  C_{n_a-1,n_b,m-1}+
      \\ \\ \displaystyle
+z \sqrt{M\omega} \sqrt{n_b+1}\sqrt{m(m-1)-\lambda}\,  C_{n_a,n_b+1,m+1} +
      \bar{z}  \sqrt{M\omega}\sqrt{n_b}\sqrt{(m+1)m-\lambda}\,  C_{n_a,n_b-1,m-1}   =0\,,
  \end{array}
\end{equation}
where $m=j+n_a-n_b$.

Notice that, for $z=0$, this recurrence gives immediately the usual harmonic oscillator levels,
\begin{equation}\label{chequeo}
    C_{n,m} \left[\omega (n_a+n_b+1)- E \right]=0 \quad \Rightarrow \quad E= \omega \left( n_a + n_b+1\right)\,.
\end{equation}

\medskip
It is not evident how to get an exact solution of this recurrence for $z\neq0$.
A difference with the Landau problem studied in \cite{Falomir-yo} is that $\left[H,b^\dagger b\right]\neq0$. A direct consequence of this, is that the eigenvalues problem does not reduce to a matricial one for  unitary \emph{irrep}'s in the discrete classes.

\subsection{The spectrum in perturbation theory}

\subsubsection{Small $|z|$}

In order to explain the structure of this spectrum we will use perturbation theory for small values of the noncommutativity parameters. For convenience, we take as unperturbed Hamiltonian $H_0$ and perturbation $V$ the operators given by
\begin{equation}\label{S21}
    \begin{array}{c} \displaystyle
      2MH_0=2M\omega\left[ a^\dagger a + b^\dagger b +1 \right]+2M\kappa s_0+\bar{z}z\left( {s_0}^2-{\mathbf{s}}^2\right)\,,
     \\ \\ \displaystyle
     2 M V= \sqrt{M\omega}\left(z a^\dagger s_+ + \overline{z}a s_- + z b^{\dagger}s_-+\overline{z} b s_+\right)\,.
    \end{array}
\end{equation}

Since $H_0$ commutes with $L_0$ and $s_0$, the unperturbed eigenvectors and eigenvalues are given by
\begin{equation}\label{S22}
   \begin{array}{c}\displaystyle
      \Psi_{n_a,n_b ,m}=\left| n_a, n_b  \right\rangle \otimes \left| \lambda , m \right\rangle
      \,, \quad H_0   \Psi_{n_a,n_b ,m} = E_{n_a,n_b,m}^{(0)} \Psi_{n_a,n_b ,m}\,,
      \\ \\ \displaystyle
     E_{n_a,n_b,m}^{(0)} =\omega\left(n_a+n_b+1\right)+\kappa m
     +\frac{\bar{z}z}{2M}\left( {m}^2-\lambda\right) \,.
   \end{array}
\end{equation}

The first order corrections to the eigenvalues in perturbation theory are simply given by
\begin{equation}\label{S23}
    E_{n_a,n_b,m}^{(1)} = \left(\Psi_{n_a,n_b ,m} ,V \Psi_{n_a,n_b ,m}\right)=0\,,
\end{equation}
and are all vanishing.

Computing second order corrections witch are given by
\begin{equation}\label{S24}
    E_{n_a,n_b,m}^{(2)} = {\sum_{n_a',n_b',m'}}'\frac{\left| \left(\Psi_{n_a',n_b',m'} ,V \Psi_{n_a,n_b,m}\right)\right|^2}{E_{n_a,n_b,m}^{(0)}-E_{n_a',n_b',m'}^{(0)}}\,,
\end{equation}
where the term with $n_a'=n_a$, $n_b'=n_b$ and $m'=m$ is excluded from the series. From (\ref{S21}) we get
\begin{equation}\label{S25}
     \begin{array}{c} \displaystyle
       \left(\Psi_{n_a',n_b',m'} ,2MV \Psi_{n_a,n_b,m}\right)=
       \\ \\ \displaystyle
=\sqrt{M\omega} z \sqrt{n_a+1} \sqrt{m(m+1)-\lambda}\, \delta_{n_a',n_a+1}\delta_{n_b',n_b}\delta_{m',m+1}        + \sqrt{M\omega} \bar{z} \sqrt{n_a} \sqrt{m(m-1)-\lambda}\, \delta_{n_a',n_a-1}\delta_{n_b',n_b}\delta_{m',m-1} +
       \\ \\ \displaystyle
+ \sqrt{M\omega} z \sqrt{n_b+1} \sqrt{m(m-1)-\lambda}\, \delta_{n_a',n_a}\delta_{n_b',n_b+1}\delta_{m',m-1}        + \sqrt{M\omega} \bar{z} \sqrt{n_b} \sqrt{m(m+1)-\lambda}\, \delta_{n_a',n_a}\delta_{n_b',n_b-1}\delta_{m',m+1}\
       \,,
     \end{array}
\end{equation}
from which it follows that
\begin{equation}\label{S26}
   \begin{array}{c} \displaystyle
      E_{n_a,l,m}^{(2)} =
     - \frac{|z|^2}{2M} \left(m^2-\lambda\right)- \frac{|z|^2}{2M} m l+ O\left(|z|^3\right) \,.
   \end{array}
\end{equation}

Then, up to second order in $|z|$, we get for the eigenvalues
\begin{equation}\label{S27}
   \begin{array}{c} \displaystyle
      E_{n_a,n_b,m}= \omega \left(n_a+n_b+1\right)+\kappa m - \frac{|z|^2}{2M} m l+ O\left(|z|^3\right)\,,
   \end{array}
\end{equation}
for any unitary \emph{irrep} of $SL(2,\mathbb{R})$.  Notice that the linear in $\kappa$ correction produces, for each given $m$, a rigid shift of the (zero order)  harmonic oscillator levels. Moreover, the term quadratic in $|z|^2$ shows a coupling with the angular momentum, thus breaking the usual degeneracy in $l$ of the isotropic harmonic oscillator spectrum, this is a difference with the previous case (the Landau problem) where the second order corrections did not break the usual degeneracy in the angular momentum of the system. Also notice that the dominant term in the $\theta$ parameter is quadratic in agreement with the Landau problem.

\subsubsection{Large $|z|$}

We will also consider the large NC parameters limit in perturbation theory. So, we now take as unperturbed Hamiltonian the operator
\begin{equation}\label{L1}
    \mathcal{H}_0:=\frac{\bar{z}z}{2M}\left( {s_0}^2-{\mathbf{s}}^2\right)+\kappa s_0+\omega\left[ a^\dagger a + b^\dagger b +1 \right]\,,
\end{equation}
and as perturbation
\begin{equation}\label{L2}
    \mathcal{V}:=\sqrt{\frac{\omega}{M}}\left(z a^\dagger s_+ + \overline{z}a s_- + z b^{\dagger}s_-+\overline{z} b s_+\right)\,.
\end{equation}
The eigenvectors and eigenvalues of $\mathcal{H}_0$ are given by
\begin{equation}\label{L3}
    \begin{array}{c}\displaystyle
    \chi_{n_a,n_b,m}:= \left|n_a, n_b \right\rangle \otimes \left|\lambda, m \right\rangle\,,
      \\\\ \displaystyle
      \mathcal{E}_{n_a,n_b,m}^{(0)}=  \frac{\bar{z}z}{2M}\left( {m}^2 -  \lambda\right)+\kappa m + \omega (n_a+n_b+1)\,.
    \end{array}
\end{equation}

The first order correction to the eigenvalues in perturbation theory vanish,
\begin{equation}\label{L4}
    \left( \chi_{n_a,n_b,m}, \mathcal{V}  \chi_{n_a,n_b,m} \right) = 0\,,
\end{equation}
while at second order $\mathcal{V}$ contributes with an $O\left(\frac{\omega}{M}\right)$ correction. Therefore, we can write
\begin{equation}\label{L5}
    \mathcal{E}_{n_a,n_b,m}=  \left\{ \frac{\bar{z}z}{2M}\left( {m}^2 -  \lambda\right)+\kappa m + \omega (n_a+n_b+1)\right\}  \left(1+O\left(\frac{\omega}{M}\right) \right)\,.
\end{equation}
Then one sees that, in the large mass limit,  the noncommutativity parameters appear as a typical energy scale for the separation of successive series of isotropic harmonic oscillator levels. For $|z|/M \gg 1$, only the states with the minimum value of $m^2$ will manifest al low energies. This conclusions are in completely agreement with the Landau problem, where we obtained the same results.


\section{Dirac Harmonic Oscillator}\label{Dirac}
The free Dirac equation in 2+1-dimensions is
\begin{equation}\label{dd1}
    \left( \imath \gamma^\mu \partial_\mu -M \right)\Psi=0\,,
\end{equation}
where we take
\begin{equation}\label{dd2}
    \gamma^0=\sigma_3\,, \quad \gamma^1=-\imath\sigma_2\,, \quad \gamma^2=\imath\sigma_1\,,
\end{equation}
which satisfy $\left[ \gamma^\mu , \gamma^\nu \right]=2 g^{\mu\nu}$ with $\left(g^{\mu\nu}\right)={\rm diag}\left(1,-1,-1\right)$. From (\ref{dd1}) we get the Hamiltoniano $H=\alpha_i\p_i+M \beta$, where  $\alpha_1=-\sigma_1$, $\alpha_2=-\sigma_2$, $\beta=\sigma_3$ and $M>0$ the mass of the particle.

In \cite{Moshi} Moshinsky and Szczepaniak  proposed to add a linear in the coordinates term, interpreting the resulting system as a \emph{Dirac oscillator} since, in the nonrelativistic limit, it reduces to a harmonic oscillator with a spin-orbit interaction. In this way, the Hamiltonian operator becomes
\begin{equation}\label{dd3}
    H=\alpha_i\left(p_i-\imath \omega \beta x_i\right) + M \beta\,,
\end{equation}
for some constant $\omega>0$.
Using the fact that $\alpha_i\beta=\imath\epsilon_{ij}\alpha_j$ is easy to see that last equation turns to be equivalent to the one of Landau problem if we identify $\omega \leftrightarrow eB/2$. Therefore, the results obtained in \cite{Falomir-yo} for the Landau problem for Dirac particles in this noncommutative space apply also to the extension of the Dirac oscillator presented here. We will not reproduce these results here and refer the reader to Section IV in \cite{Falomir-yo}.

\section{Conclusions}\label{conclusion}


In this article we have considered models of Schr\"{o}dinger and Dirac oscillators in a space-time with a kind of noncommutativity, both in coordinates and momenta, induced by deforming the canonical commutators by terms proportional to the generators in a unitary irreducible representation of the Lorentz group in the 2+1-dimensional Minkowski space, isomorphic to $SL(2,\mathbb{R})/\mathbb{Z}_2$. Since this is a noncompact Lie group, its unitary \emph{irrep}'s are not of finite dimension. This noncommutative phase space has been previously considered in \cite{Falomir-yo}.

In order to get a full view about the realization problem of this algebra we analyze Levi decomposition and we obtain that it can be represented as a direct sum of the usual commutative operators plus the generators of $SL(2,R)$. This realization is not unique, they can differ by a canonical transformation of the dynamical variables.

We realize the algebra by means of a shift in the canonical coordinates and momenta with terms proportional to the generators of $SL(2,\mathbb{R})/\mathbb{Z}_2$ in a unitary \emph{irrep}. In particular, the shift in momenta can also be interpreted as the introduction of a non-Abelian magnetic field.

Consequently, the number of dynamical variables is enlarged and the Hilbert space gets the structure of a direct product, one factor for the state vectors of the ordinary system in the normal space and the other for the component of the state vectors in the representation space of this \emph{irrep} of $SL(2,\mathbb{R})$.

It was shown in \cite{Falomir-yo} that total generators of the Lorentz transformations can be constructed which correctly transform all the operators, thus realizing the Lie algebra $sl(2,\mathbb{R})$ on the Hilbert space of these quantum-mechanical system.

In this framework, we have considered the modified  Hamiltonians of the Harmonic and Dirac Oscillators. We have analyzed these models for both discrete and continuous classes of \emph{irrep}'s of $SL(2,\mathbb{R})$, getting  linear infinite recursion relations for the coefficients in the development of the Hamiltonian eigenvectors in terms of conveniently chosen bases of the Hilbert space. The spectrum of these models have been studied also in perturbation theory, both for small and large noncommutativity parameters $z= \theta M\omega+\imath \kappa$.

For small $z$, we have shown that the spectrum of the extension of the harmonic oscillator is the one of the isotropic two-dimensional oscillator rigidly shifted by  $\kappa m$,  a term proportional to the eigenvalue of $s_0$. Moreover, second order corrections in $|z|$ break the degeneracy in the angular momentum. In particular,  the dominant correction in the $\theta$ parameter is quadratic. In the large $|z|$ limit,  the noncommutativity parameters appear as a typical energy scale for the separation of successive series of isotropic harmonic oscillator levels.

On the other hand, we have shown that the extension of the Dirac oscillator to this noncommutative space is completely equivalent to the extension of the Landau problem for Dirac particles studied in \cite{Falomir-yo}.

Let us mention that, contrary to the case of conventional NC Quantum Mechanics, we find no constraint between the parameters referring to no-commutativity in coordinates and momenta. Rather, both $\kappa$ and $\theta$ play a similar role.

As discussed in \cite{Falomir-yo}, these models do not correspond to a smooth deformation of the commutative ones. Rather, the Hilbert space takes the structure a of the direct product of the usual one times the representation space of an \emph{irrep} of $SL(2,\mathbb{R})$. Since these \emph{irrep}'s  can be explicitly realized in terms of spaces of square-integrable functions (functions defined on the unit circle for the continuous classes of  \emph{irrep}'s and analytic functions on the unit open disk for the discrete classes of  \emph{irrep}'s - see \cite{Bargmann,Falomir-yo}) the examples studied in this article can also be considered as equivalent to models of quantum mechanical particles living in a space with an additional compact dimension, with $|z|$ playing the role of the inverse of a typical length.


\vskip 1cm

\noindent
\textbf{Acknowledgements}: F.V. acknowlege finantial support from CONICET, Argentina. This work was partially supported by Universidad Nacional de La Plata (Proy.~11/X615), Argentina.



\end{document}